\documentclass[aps,prl,superscriptaddress,12pt]{revtex4}

\usepackage{times}

\usepackage{amsmath}
\usepackage{amsfonts}
\usepackage{amssymb}
\usepackage{epsfig}

\begin{document}
\normalsize{\textbf{Science, Vol.316, p.1169, 2007}}\\\\

\title{An On-Demand Coherent Single Electron Source}

\author
{G. F{\`e}ve,$^{1}$ A. Mah\'{e},$^{1}$ J.-M. Berroir,$^{1}$ T.
Kontos,$^{1}$ B. Pla\c{c}ais,$^{1}$ D.C. Glattli,$^{1,2,\ast}$\\A.
Cavanna,$^{3}$ B. Etienne,$^{3}$
Y. Jin$^{3}$\\
\normalsize{$^{1}$Laboratoire Pierre Aigrain, D{\'e}partement de
Physique de l'Ecole Normale Sup\'erieure}\\
\normalsize{24 rue Lhomond, 75231 Paris Cedex
05, France}\\
\normalsize{$^{2}$Service de Physique de l'Etat Condens{\'e},
CEA Saclay}\\
\normalsize{F-91191 Gif-sur-Yvette, France}\\
\normalsize{$^{3}$Laboratoire de Photonique et Nanostructures, UPR20
CNRS}\\
\normalsize{Route de Nozay, 91460 Marcoussis Cedex, France}\\
\normalsize{$^\ast$ To whom correspondence should be addressed;
E-mail:  glattli@lpa.ens.fr.} }

\begin{abstract}
We report on the electron analog of the single photon gun. On demand
single electron injection in a quantum conductor was obtained using
a quantum dot connected to the conductor via a tunnel barrier.
Electron emission is triggered by application of a potential step
which compensates the dot charging energy. Depending on the barrier
transparency the quantum emission time ranges from 0.1 to 10
nanoseconds. The single electron source should prove useful for the
implementation of quantum bits in ballistic conductors. Additionally
periodic sequences of single electron emission and absorption
generate a quantized AC-current.

\end{abstract}

\maketitle

In quantum optics, a single photon source is an essential building
block for the manipulation of the smallest amount of information
coded by a quantum state: a qubit \cite{Imamoglu_94,Gisin_Rev02}.
Combined with beam-splitters, polarizers and projective measurements
several photonic qubits can be manipulated to process quantum
information \cite{Kok_Rev07}. The most celebrated case is the
secured transmission of the information using quantum cryptography.
Similarly, one expects that electrons propagating ballistically in
ultra-pure low dimensional conductors can realize quantum logic
tasks in perfect analogy with photons propagating in optical media
\cite{flying_quBits1,flying_quBits2,flying_quBits3}. The analogy has
a long history \cite{Analogy} and has provided illuminating
comparisons between the intensity of light and that of electrical
current, between photon noise and electrical shot noise
\cite{Butt92,Blanter00} and more recently between photon and
electron quantum entanglement \cite{Beenakker_03,
Samuelson_04,Beenakker_06}. Interestingly, electrons being Fermions,
entanglement offers new routes not possible with photons
\cite{Beenakker_06}. Practically, electronic analogs of
beam-splitters, Fabry-P\'erot and Mach-Zehnder interferometers
\cite{VanHouten_Beenakker,Heiblum_03} have been realized in
ballistic conductors providing the necessary quantum gate for an
'all linear' electron optics quantum computation. Yet missing were
the single electron source and the single electron detector
\cite{note1} suitable for coherent emission and projective
measurements. The former initializes quantum states, while the
latter reads the final states after electrons have passed through
the quantum gates.

Unlike the case of photons, realization of single electron sources
is expected to be simpler because of Fermi statistics and Coulomb
interaction. For example, considering a voltage biased single mode
conductor, a contact at energy $eV$ above the energy of the other
contact is known to inject single electrons into the conductor at a
regular rate $eV/h$, thereby leading to quantization of the dc
current in Quantum Point Contacts \cite{VanWees_88a,VanWees_88b}. A
second example is the electron pump where a dc current is produced
by sequential time-controlled transfer of single electrons between
metallic islands in series \cite{pump1,pump2} or manipulation of
tunnel barriers of quantum dots \cite{pump3,pump4}. The cost in
Coulomb charging energy to add or remove an electron ensures a well
defined electron number in each island or dot. These two sources are
however not useful for quantum information. In the first case, there
is no time control of the electron injection. As only statistical
measurements are possible, the biased contact is suitable for
demonstrating coherent phenomena such as interferences or electron
entanglement \cite{Beenakker_03,Samuelson_04} but not for
manipulating quantum information. In the second example, time
controlled injection can be realized, but the energy of emitted
electrons is expected to spread, at random, in an energy range much
larger than the tunneling rate (typically a fraction of the charging
energy, depending on the pumping conditions). The statistical
distribution in energy will smear coherent effects required for
manipulating the quantum information. Finally, a third approach has
been theoretically proposed in Refs.
\cite{Levitov96,Jonkheere05,Levitov05} considering voltages pulses
$V(t)$ applied to an ohmic contact. When the Faraday flux $e\int
V(t')dt'/h$ is an integer, an integer number of electrons is
injected. Here noiseless injection requires to have a special
Lorenzian shape of the pulse and exact integer value otherwise
logarithmic divergenge of the charge fluctuations occurs. No
experiment is available yet to test these ideas.

We report on the realization of a time controlled single electron
source suitable for coherent manipulation of ballistic electronic
qubits which emits the electrons into a well defined quantum state.
The injection scheme is different from those considered above. The
source is made of a quantum dot, realized in a 2D electron gas in
GaAs semiconductors, and tunnel-coupled to the conductor through a
quantum point contact (QPC). By applying a sudden voltage step on a
capacitively coupled gate, the charging energy is compensated and
the electron occupying the highest energy level of the dot is
emitted. The final state of the electron is a coherent wave-packet
propagating away in the conductor. Its energy width is given by the
inverse tunneling time, as required for on-demand single particle
source, and independent on temperature. Its mean energy can be
adjusted above the Fermi energy by tuning the voltage step
amplitude.  The circuit (Fig.1A), is realized in a 2D electron gas
(2DEG) in a GaAsAl/GaAs heterojunction of nominal density
$n_{s}=1.7\times 10^{15}$ $m^{-2}$ and mobility $\mu=260$
$V^{-1}m^{2}s^{-1}$. The dot is electrostatically coupled to a
metallic top gate, 100nm above the 2DEG, whose ac voltage,
$V_{exc}$, controls the dot potential at the subnanosecond
timescale. For all measurements, the electronic temperature is about
200 mK  and a magnetic field $B\approx 1.3\;\rm{T}$  is applied to
the sample so as to work in the quantum Hall regime with no spin
degeneracy. The QPC dc gate voltage $V_{G}$ is tuned to control the
transmission $D$ of a single edge state as well as the dc dot
potential. As reported \cite{Gabelli06Science}, this circuit
constitutes the paradigm of a quantum coherent RC circuit where
coherence is seen to strongly affect the charge relaxation dynamics.
From this study, the charging energy $\Delta + e^{2}/C \approx
\Delta \approx 2.5$ K was extracted \cite{note2}. Here the large top
gate capacitance makes the Coulomb energy $e^{2}/C$ unusually small
and the total charging energy identifies to the energy level spacing
$\Delta$.

In Ref.\cite{Gabelli06Science}, the linear response of the current
to the ac top gate voltage was investigated and the ac charge
amplitude was much lower than the elementary charge $e$. Here, in
order to achieve single charge injection we have to apply a high
amplitude excitation ($V_{exc} \sim \Delta/e$) and go beyond the
linear regime. When an electron is suddenly brought above the Fermi
energy of the lead, it is expected to escape the dot at a typical
tunnel rate $\tau^{-1}= D\Delta/h$, where $\Delta/h$ is the attempt
frequency and $D$ the transmission probability. This gives
nanosecond timescales for which single charge detection is still out
of reach experimentally. To increase the signal to noise ratio, a
statistical average over many individual events is used by
generating repetitive sequences of single electron emission followed
by single electron absorption (or hole emission) as sketched in
Fig.1A. This is realized by applying a periodic square wave voltage
amplitude $\approx \Delta/e$ to the top gate. Fig.1B shows typical
temporal traces of the current averaged over few seconds for a
repetition period of $\mathcal{T}=32\,ns$. The single electron
events remarkably reconstruct the exponential current decay of an RC
circuit. When decreasing  transmission  $D$ from $\approx 0.03$ to
$\approx 0.002$, the relaxation time $\tau$, extracted from the
exponential decay, increases from $0.9$ ns to $10$ ns. For the two
highest transmissions in Fig.1B, $\tau \ll \mathcal{T}/2$, the
current decays to zero and the mean transferred charge per half
period is constant. For the smallest transmission, $\tau \sim
\mathcal{T}/2$, the mean emitted charge decreases as electrons have
reduced probability to escape the dot. These time-domain
measurements are limited by the $1$ GHz bandwidth of the acquisition
card and give access to the few nanoseconds injection times
corresponding to small transmissions $D \lesssim 0.03$.

In order to get a  better understanding of the above results, we
extend the harmonic linear response theory of a quantum RC circuit
\cite{BPT93PL,BPT93PRL,PTB96PRB} to calculate the non-linear
response to a high amplitude square excitation voltage ($eV_{exc}\gg
hf$). Calculation shows that the circuit still behaves as an RC
circuit with a current given by:
\begin{eqnarray}
I(t) &=& \frac{q}{\tau} e^{-t/\tau} \quad \textrm{for}\quad
0 \leq \tau \leq \mathcal{T}/2 \label{I} \\
q &=&  e \int d\epsilon
N(\epsilon)[f(\epsilon-2eV_{exc})-f(\epsilon)] \label{q} \\
\tau & = &\frac{h}{2} \frac{ \int d\epsilon
N(\epsilon)^{2}[f(\epsilon-2eV_{exc})-f(\epsilon)] }{\int d\epsilon
N(\epsilon)[f(\epsilon-2eV_{exc})-f(\epsilon)]}\label{Rq}\label{tau}
\end{eqnarray}
where $N(\epsilon)$ is the dot density of states and $f(\epsilon)$
denotes the Fermi-Dirac distribution. The non-linear capacitance and
charge relaxation resistance can be defined respectively by
$\widetilde{C_{q}}\equiv q/2V_{exc}$ and
$\widetilde{R_{q}}\equiv\tau/\widetilde{C_{q}}$. For unit
transmission $D=1$, electrons are fully delocalized, $N(\epsilon)$
is uniform and the charge $q$ evolves linearly with $V_{exc}$ as
expected. At the opposite, for low transmission, $N(\epsilon)$ is
sharply peaked on well resolved energy levels, and $q$ exhibits a
staircase dependence on $V_{exc}$ with steep steps whenever one
electronic level is brought above the Fermi energy. Thus our
calculations establish the sketch of single electron injection
depicted in Fig.1. For a dot energy spectrum with constant level
spacing $\Delta$, a remarkable situation occurs when
$2eV_{exc}=\Delta$, as $q=e$ and $\widetilde{C_{q}} = e^{2}/\Delta$
irrespective of the transmission $D$ and of the dc dot potential. As
a matter of fact, Eq.\ref{q} shows that, in these conditions,  $q$
is given by integrating $N(\epsilon)$ over exactly one level
spacing. For $D <<1$, we recover the Landauer formula for the
resistance, $\widetilde{R_{q}}=\frac{h}{D e^{2}}$ and the escape
time is given by $\tau=h/D\Delta$, as expected from a semiclassical
approach. The exponential current decay, the constant injection
charge for $\tau \ll \mathcal{T}/2$, as well as the decrease of
$\tau$ with transmission $D$, account well for our experimental
observations in Fig.1B.

For a more accurate experimental determination of $q$ and $\tau$ and
to investigate subnanosecond time scales, we consider in the
following measurements of the current first harmonic, $I_{\omega}$ ,
at higher frequencies $f =\omega/2\pi=1/\mathcal{T}$.  As a matter
of fact, following Eq.\ref{I}, we have:
\begin{eqnarray}
I_{\omega}&=&\frac{2qf}{1-i\omega \tau}
\end{eqnarray}
so that the modulus $|I_{\omega}|$ and the phase $\phi$
($\tan(\phi)=\omega\tau$) allow for the determination of $q$ and
$\tau$.

Fig.2A shows $|I_{\omega}|$  measured as a function of QPC gate
voltage $V_{G}$ at $f=180$ MHz for increasing values of the
excitation voltage $2eV_{exc}$. The range of $V_{G}$ maps the full
transmission excursion $D=0$-$1$. The low excitation
$2eV_{exc}=\Delta/4$ data nearly correspond to the linear response
reported in Ref.\cite{Gabelli06Science}. The current exhibits strong
oscillations reflecting the variation with $V_{G}$ of the dot
density of states at the Fermi energy. At larger excitation
voltages, the current peaks are broadened as expected from
Eq.\ref{q} when $2eV_{exc}$ gets larger than $k_{B}T$. For
$2eV_{exc}=\Delta$, the oscillations disappear completely and
$|I_{\omega}|=2ef$, down to a low transmission threshold $D\sim
0.05$. The oscillations reappear for larger excitations. The
constant current $|I_{\omega}|=2ef$ is the frequency-domain
counterpart of the constant charge regime observed in the
time-domain, for the injection/absorption of a single electron per
half period. The cut-off observed for $D\lesssim 0.02$ corresponds
to the limit $\omega\tau\gtrsim 1$ where the escape time $\tau$
exceeds $\mathcal{T}/2$. The constant $\widetilde{C_{q}}$ regime
obtained for $2eV_{exc}=\Delta$ can be viewed on a striking manner
in a Nyquist representation of Fig.2B. The corresponding diagram is
the half-circle characteristic of an RC circuit with a constant
capacitance $e^{2}/\Delta$ and transmission dependent resistance. By
contrast the curves obtained for larger or smaller excitations
exhibit strong capacitance oscillations.

Fig.2C represents the phase $\phi=\arctan(\omega\tau)$ of the
current as a function of $V_{G}$ for different excitation voltages.
$\phi$ shows a quasi monotonic $\pi/2$ sweep in increasing
transmission. The absence of significant oscillations proves that
$\tau$ is nearly insensitive to the dot potential. As seen in the
figure, $\tau$ is also independent of $V_{exc}$. In Fig.3, we have
gathered the values of $\tau(V_{G})$ obtained from 1GHz bandwidth
time-domain measurements at 31.25 MHz repetition rate and from
frequency-domain measurements at 180 and 515 MHz. The whole
measurements probe a very broad transmission range ($D=0.002-0.2$)
corresponding to escape times varying from 10 ns to 100 ps. In the
overlapping range, the different independent determinations coincide
within error bars, agreeing quantitatively with the prediction
$\tau= h/D\Delta$ also represented in Fig.3, where the dependence
$D(V_{G})$ is deduced from the linear regime
\cite{Gabelli06Science}.

We now discuss the conditions for single electron injection leading
to a good quantization of the ac current as a figure of merit of
single charge injection. Fig.4A represents $|I_{\omega}|$ as
function of $V_{exc}$ for typical values of the dc dot potential at
fixed transmissions $D\approx 0.2$ and $D \approx 0.9$. Transmission
$D\approx 0.2$ is low enough for the electronic states to be well
resolved, as sketched in the inset of Fig.4A (left), but still large
for the escape time to be shorter than $\mathcal{T}/2$.  When the
Fermi energy lies exactly in the middle of a density of states
valley, we observe a well pronounced $|I_{\omega}|=2ef$ current
plateau centered on $2eV_{exc}=\Delta$. Whereas the current plateau
resolution is noise limited to better than $1\%$ (for a 10 seconds
acquisition time), the plateau value is determined with an
uncertainty of $5\%$ due to systematic calibration error.  We note
that at this working point the plateau is robust upon variation of
the parameters. By contrast, if the Fermi energy lies on a peak,
there is still a current plateau but its value is arbitrary and very
sensitive to parameter variations. These two working points
illustrate the importance of having a well defined charge in the dot
prior to injection. In the first case the charge is well defined and
suitable for charge injection. In the second case the equilibrium
dot charge fluctuates. In particular, when the energy level is
exactly resonant with the Fermi energy, its mean occupation at
equilibrium is $1/2$ and the measured value of the plateau is
$1/2\times 2ef=ef$ (see Fig.4A (left)). Thus, this working point is
not suitable for a single electron source. Upon increasing
transmission, even for a suitable working point, the dot charge
quantization can be lost because of quantum fluctuations. First, the
width of the ac current plateaus reduces and finally nearly vanishes
for $D\approx 0.9$. Note that for different transmissions, all
curves cross at $|I_{\omega}|=2ef$ for $2eV_{exc}=\Delta$ reflecting
the constant value of $\widetilde{C_{q}}$ discussed above. Finally,
domains of good charge quantization are best shown on the
two-dimensional color plot of Fig.4.B-upper where the modulus of the
current is represented in color scale. The vertical axis stands for
the excitation voltage $V_{exc}$ and the horizontal axis for the
gate voltage $V_{G}$. The white diamonds correspond to large domains
of constant current $|I_{\omega}|=2ef$ suitable for single electron
injection. At high transmissions the diamonds are blurred by dot
charge fluctuations as discussed previously. On the opposite, for
small transmissions, even when the dot charge quantization is good,
current quantization is lost because of long escape time
$\omega\tau>>1$,  and the current goes to zero. At $180 MHz$,
optimal working conditions are obtained for $D\approx 0.2$.
Experimental results of Fig.4 are compared with our theoretical
model (Eqs.\ref{q} and \ref{Rq}) without any adjustable parameter
(solid lines in Fig.4a and lower plot in Fig.4b)\footnote{We use the
1D modeling of our circuit (density of states, transmission,
dot-gates coupling) described in reference \cite{Gabelli06Science}}.
The agreement between measurements and theoretical predictions is
excellent which shows that our single electron source lends itself
to quantitative modeling.

The availability of a coherent source of single electrons emitted on
demand from a single energy level on nanosecond time scale opens the
way for a new generation of experiments never possible before.
Synchronization of similar sources could be used in the future to
probe electron anti-bunching, electron entanglement in multi-lead
conductors or to generate electronic flying qubits in ballistic
conductors.\\
The LPA is the CNRS-ENS UMR8551 associated with universities Paris 6
and Paris 7. The research has been supported by SESAME Ile-de-France
and ANR-05-NANO-028 contracts.

\clearpage

\begin{figure}[hhhhhh]
\centerline{\includegraphics[width=15 cm,
keepaspectratio]{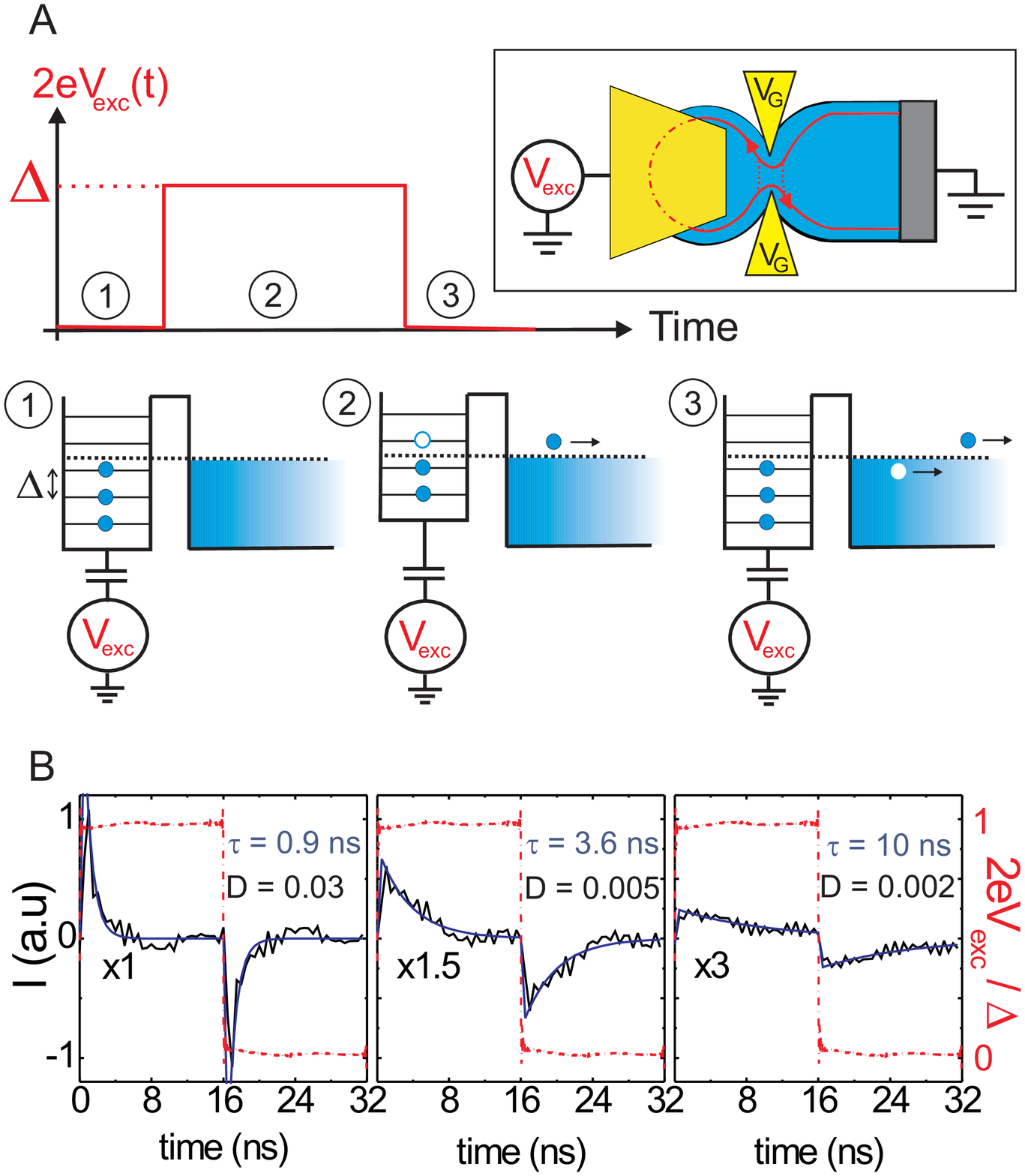}} \caption{Single charge injection. A)
Schematic of single charge injection. Starting from an antiresonant
situation where the Fermi energy lies between two energy levels of
the dot (step 1), the dot potential is increased by $\Delta$
bringing one occupied level above the Fermi energy (step 2). One
electron then escapes the dot on the mean time
$\tau=\frac{h}{D\Delta}$. The dot potential is then brought back to
its initial value (step 3) where one electron can enter it, leaving
a hole in the Fermi sea. Inset: The quantum RC circuit : one edge
channel is transmitted inside the submicrometer dot with
transmission $D$ tuned by the QPC gate voltage $V_{G}$. The dot
potential is varied by a radiofrequency excitation $V_{exc}$ applied
on a macroscopic gate located on top of the dot. The electrostatic
potential can also be tuned by $V_{G}$ due to the electrostatic
coupling between the dot and the QPC. B) Time-domain measurement of
the average current (black curves) on one period of the excitation
signal (red curves) at $2eV_{exc}=\Delta$ for three values of the
transmission $D$. The relaxation time $\tau$ is deduced from an
exponential fit (blue curve).}
\end{figure}

\newpage

\begin{figure}[hhhhhh]
\centerline{\includegraphics[width=15 cm,
keepaspectratio]{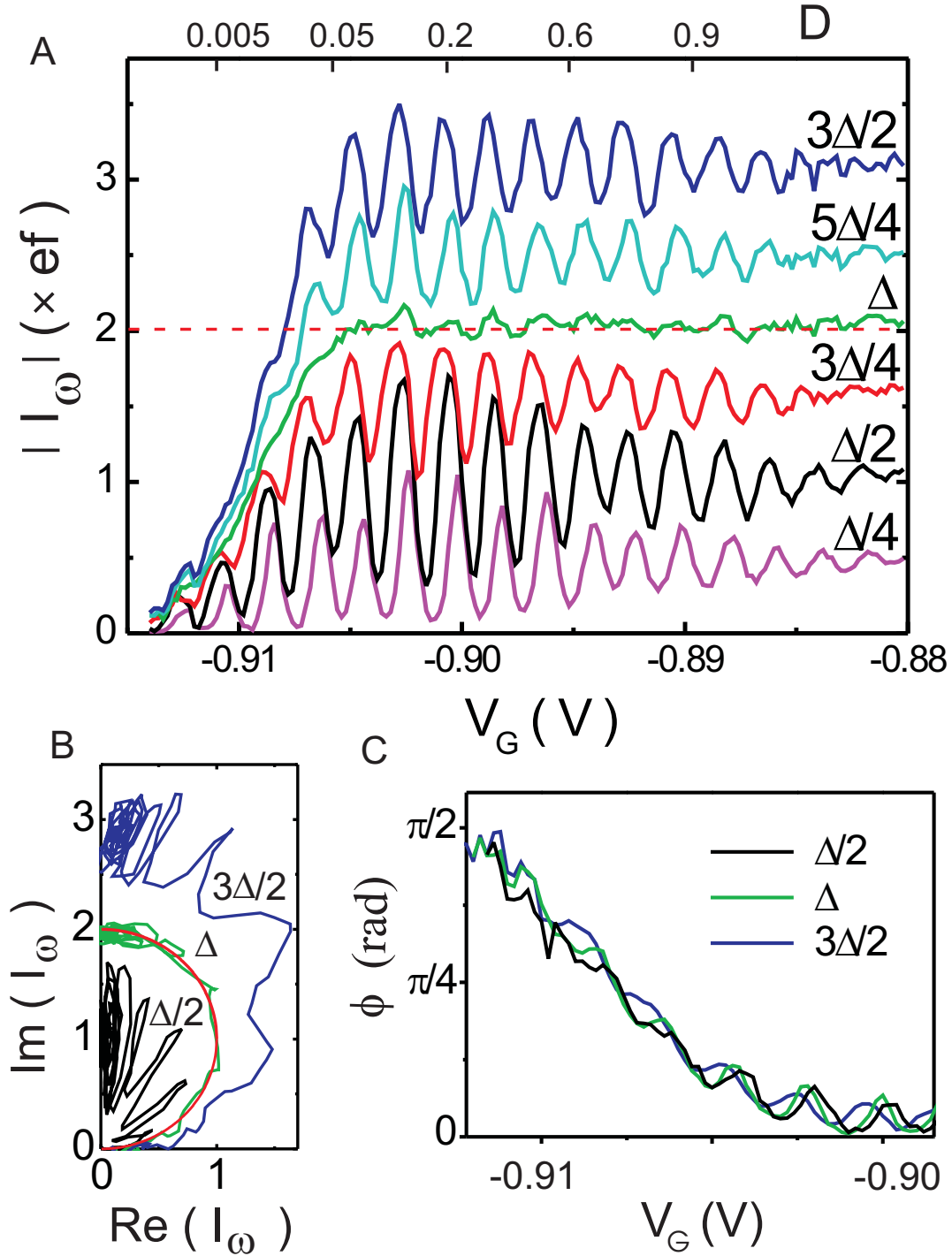}} \caption{ $I_{\omega}$ as a function of
$V_{G}$ at $f=180$ MHz for different values of the excitation
amplitude $2eV_{exc}$. Transmission $D$ is also indicated. A)
Modulus $|I_{\omega}|$. The dashed line is the constant value
$|I_{\omega}|=2ef$. B) Nyquist representation ($Im(I_{\omega})$ vs
$Re(I_{\omega})$). The red curve corresponds to an RC circuit of
constant capacitance $e^{2}/\Delta$ and varying resistance. C) Phase
$\phi$. The phase $\phi$ is independent of $V_{exc}$ .}
\end{figure}
\newpage

\begin{figure}[hhhhhh]
\centerline{\includegraphics[width=15 cm,
keepaspectratio]{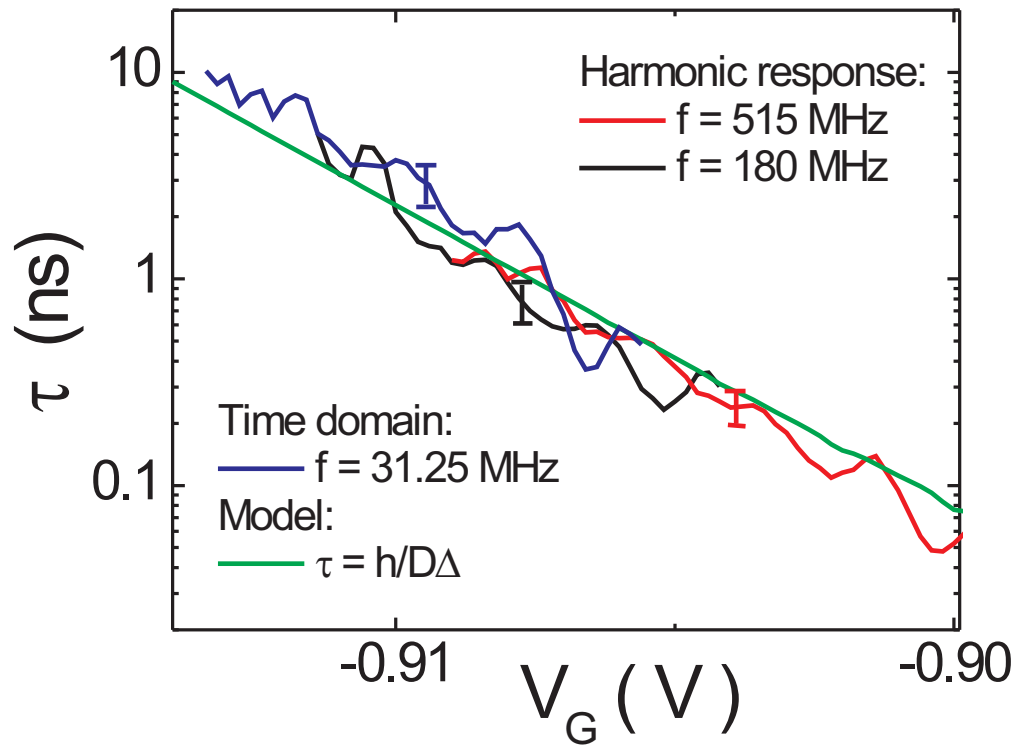}} \caption {Escape time $\tau$ in
logarithmic scale as a function of QPC gate voltage $V_{G}$ :
experiments and model.}
\end{figure}

\newpage

\begin{figure}[hhhhhh]
\centerline{\includegraphics[width=15 cm,
keepaspectratio]{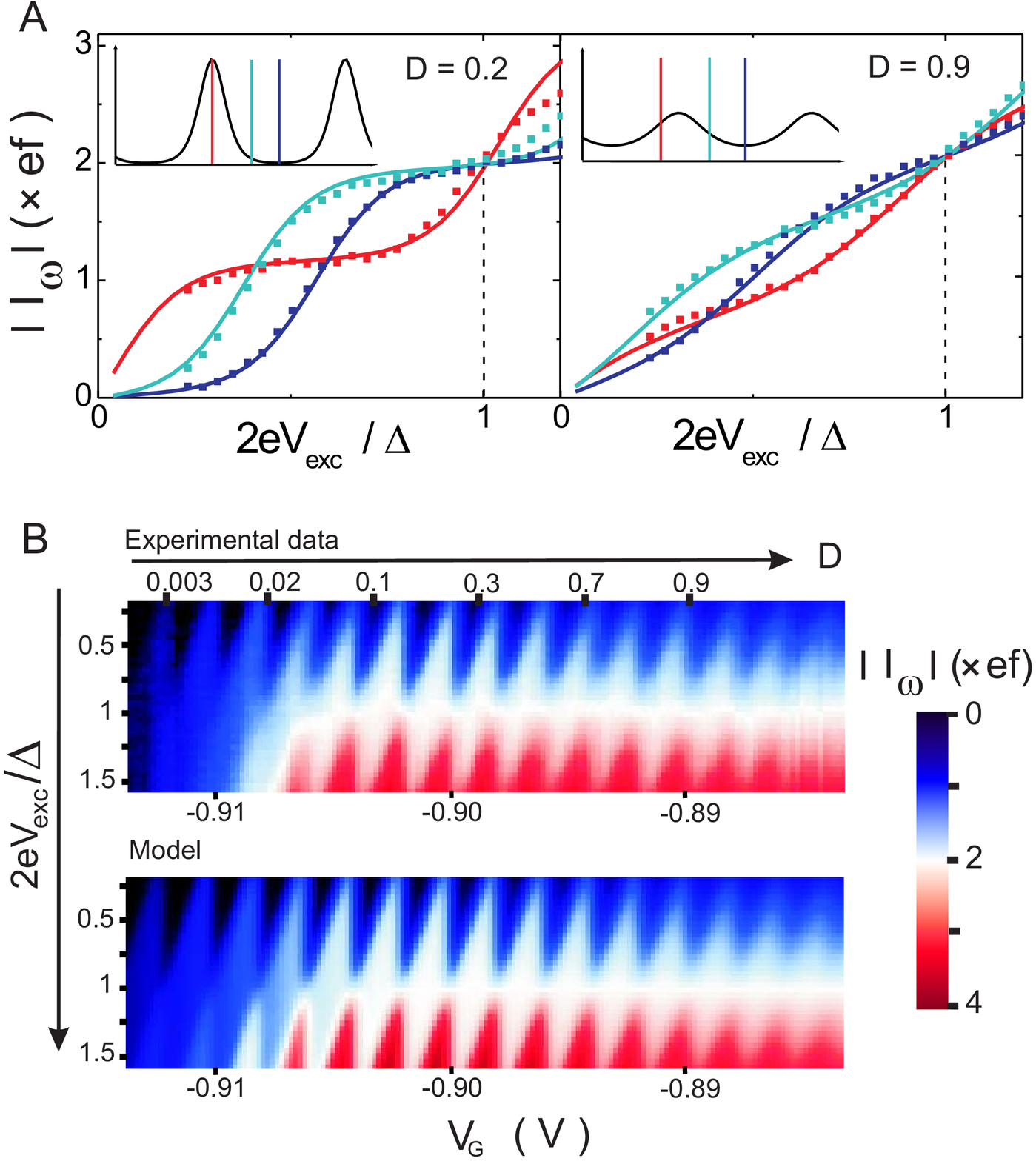}} \caption{Quantization of the ac current.
A) $|I_{\omega}|$ as a function of $2eV_{exc}/\Delta$ for different
 dot potentials at $D\approx 0.2$ (left) and $D\approx 0.9$ (right).
 Points correspond to experimental values and lines to theoretical
 predictions. Insets: schematic representation of the dot density of
 states $N(\epsilon)$. The color bars indicate the dot potential for
 the corresponding experimental data. B) Color plot of $|I_{\omega}|$
 as a function of $2eV_{exc}/\Delta$ and $V_{G}$: experiments (upper)
 and model (lower).}
\end{figure}

\newpage


\begin{thebibliography}{9}
\bibitem{note1} single electrons detectors have been realized but
are yet too slow compare with the electron transit time in the
measurement leads.
\end{thebibliography}
\end{document}